\documentclass[manuscript]{aastex63}

\usepackage{multirow}
\usepackage{txfonts}
\usepackage{latexsym,bm}

\newcommand{\qinemail}{qingang@hit.edu.cn}
\newcommand{\hit}{School of Science, Harbin Institute of
Technology, Shenzhen, 518055, China}
\newcommand{\hitqin}{\hit; \qinemail}

\submitjournal{ApJ}

\turnoffedit

\shorttitle{PREDICTION OF SOLAR CYCLES 25 AND 26}
\shortauthors{WU \& QIN}

\begin{document}

\title{Predicting Sunspot Numbers for Solar Cycles 25 and 26}

\correspondingauthor{G. Qin}
\email{\qinemail}

\author[0000-0002-5776-455X]{S.-S. Wu}
\affiliation{\hitqin}

\author[0000-0002-3437-3716]{G. Qin}
\affiliation{\hitqin}

\begin{abstract}
The prediction of solar activity is important for advanced technologies and space
activities. The peak sunspot number (SSN), which can represent the solar activity,
has declined continuously in the past four solar cycles (21$-$24), and the Sun would
experience a Dalton-like minimum, or even the Maunder-like minimum, if the declining
trend continues in the following several cycles, so that the predictions of solar
activity for cycles 25 and 26 are crucial. In Qin \& Wu, 2018, ApJ, we established
an SSN prediction model denoted as two-parameter modified logistic prediction (TMLP)
model, which can predict the variation of SSNs in a solar cycle if the start time
of the cycle has been determined. In this work, we obtain a new model denoted as
TMLP-extension (TMLP-E),
\edit1{which can predict the solar cycle nearly two cycles in advance, so that
the predictions of cycles 25 and 26 are made}.
It is found that the predicted solar maximum, ascent time, and cycle length are
115.1, 4.84 yr, and 11.06 yr, respectively, for cycle 25, and 107.3, 4.80 yr,
and 10.97 yr, respectively, for cycle 26. The solar activities of cycles 25 and
26 are predicted to be at the same level as that of cycle 24, but will not
decrease further. We therefore suggest that the cycles 24$-$26 are at a minimum
of Gleissberg cycle.
\end{abstract}

\keywords{Solar cycle (1487); Sunspot cycle (1650); Sunspot number (1652); Solar activity (1475)}

\section{Introduction}

The solar activity has vital influences on the solar-terrestrial environment, which
further affects the health of human beings and the safety of spacecraft as well
as the reliability of navigation and communication
\citep[e.g.,][]{Lanzerotti17, MertensEA18, ShenAQin18, WuAQin18}. Most of solar
activities range from several minutes to decadal-scale, such as the quasi-11-year
period of solar activity strength, which is called the solar activity cycle and can
be well represented by the sunspot number (SSN)
\citep[e.g.,][]{BaloghEA14, Hathaway15, LinEA19, Petrovay20, ChenEA21}.

Long-term predictions of solar activity are essential for planning future space
missions and understanding the underlying mechanism of the solar cycle.
The predictions for solar cycles 25 and 26 are more crucial because the solar
maximum, defined as the peak monthly smoothed SSN of a solar cycle, has declined
continuously in the past 4 cycles (21$-$24). Solar cycle 24 is the second lowest
cycle with the solar maximum being 116.4 since the Dalton minimum, recorded around
the year 1810. A new Dalton-like minimum even the Maunder-like minimum may occur
if the solar activity decreases further in cycles 25 and 26, which will cause
some important space weather effects.

A lot of research has been focused on the prediction of the solar maximum of cycle 
25 in recent years. Firstly, some research suggested that the declining trend of
solar activity will continue. The solar maximum of cycle 25 is predicted to be
about 14\% \citep{Macario-RojasEA18}, 24\% \citep{LabonvilleEA19},
and 31\% \citep{SinghEA19} lower than that of cycle 24. In addition, several works
obtained much lower values, e.g., 50 $\pm$ 15 \citep{Kitiashvili20} and 57 $\pm$ 17
\citep{CovasEA19}, for the solar maximum of cycle 25, which indicates that cycle 25
would be the weakest cycle since the Maunder minimum
\citep[1645$-$1715; e.g.,][]{Eddy76}. Secondly, some other studies, however,
inferred that the declining trend of solar activity will break and cycle 25
would be stronger than cycle 24. The solar maximum of cycle 25 is forecast to be
about 16\% \citep{PesnellASchatten18}, 24\% \citep{KakadEA20}, 30\% \citep{Du20},
32\% \citep{SarpEA18}, and 45\% \citep{LiEA18} greater than that of cycle 24.
A pretty large value of 228.8 $\pm$ 40.5 was suggested by \citet{HanAYin19}.
Finally, there are also some works to predict that the solar maximum of cycle 25
would be similar to that of cycle 24 (the difference is within 10\%)
\citep[e.g.,][]{BhowmikANandy18, JiangEA18, OkohEA18, UptonAHathaway18, GoncalvesEA20, Lee20}.

Compared with the cycle 25, researchers seldom work on the predictions for the 
cycle 26 since longer term forecasts are more difficult. \citet{Charvatova09}
suggested that cycles 24$-$26 would be a repeat of cycles 11$-$13, so that the
solar maxima of cycles 24$-$26 should be 140.3, 74.6, and 87.9, respectively.
\citet{Hiremath08} forecasted the amplitude and period of cycles 24$-$38.
The solar maximum was predicted to be 110 $\pm$ 11 for both cycle 24 and cycle 25,
while cycle 26 was supposed to experience a very high solar activity. Note that,
these early predictions are made based on the Version 1.0 SSN as the Version 2.0
SSN was not released until 2015 \citep{CletteEA2015, CletteALefevre2016}.
The observed solar activity of cycle 24, however, is at a relatively low level
with the solar maximum being 81.9 for the Version 1.0 SSN. \citet{Abdusamatov07}
predicted that the solar maxima of cycles 24, 25, and 26 would be 70 $\pm$ 10,
50 $\pm$ 15, and 35 $\pm$ 20, respectively, for the Version 1.0 SSN.
\citet{SinghABhargawa19} also predicted that the declining trend of solar activity
would continue with the solar maxima of cycles 25 and 26 being 89 $\pm$ 9 and
78 $\pm$ 7, respectively, for the Version 2.0 SSN. They suggested that the Sun
would experience a Dalton-like minimum by the year 2043. In \citet{Javaraiah17},
a low value of 30$-$40 for cycles 25 and 26 was predicted, and the epochs of
cycles 25 and 26 were suggested to be at a minimum of Gleissberg cycle, which is
a 60$-$120 yr variation in solar cycle amplitude \citep[e.g.,][]{Gleissberg39, Petrovay20}.

It is shown that the solar maximum of cycle 25 predicted by various methods ranges
from a pretty low value of 50 up to a very large value of 228.8
\edit1{\citep[see also][]{Nandy21}, and}
\citet{Pesnell16} showed that 105 predictions
\edit1{based on diverse techniques}
for cycle 24 also gave a wide prediction range. Besides, predictions for cycle
26 are rare so far,
\edit1{and some early predictions may not be accurate since they have failed in
predicting cycle 24 \citep[e.g.,][]{Hiremath08, Charvatova09}.}


\edit1{Solar cycle is understood to be driven by the magnetohydrodynamic dynamo
mechanism, so that dynamo model based predictions are received with increasing
confidence although the technique is utilized to predict the solar cycle only
since cycle 24 \citep{DikpatiEA06, ChoudhuriEA07, Petrovay20, Nandy21}.
Though the physical model based predictions for cycle 24 have a large divergence,
\citet{Nandy21} pointed out that the predictions for cycle 25 have converged,
and the mean predicted amplitude by physics-based forecasts is very similar to
the observed amplitude of cycle 24. However, reasonably accurate predictions
by physical models are possible only one cycle in advance because the efficient
transport of magnetic flux reduces the dynamical memory of the sunspot cycle to
only one cycle \citep{YeatesEA08, KarakANandy12, Munoz-JaramilloEA12, Nandy21}.
Besides, the physical based predictions need some external data such as the
polar field that is accessible only in recent decades, so that these predictions
could not be checked by more solar cycles at present. Therefore, the statistical
forecast is still valuable for us to study since it may be able to predict
the solar cycle with good accuracy two cycles in advance.}

Recently, we established an SSN prediction model, denoted as two-parameter
modified logistic prediction (hereafter referred to as TMLP) model
\citep{QinAWu18}, statistically. The model can predict the variation of SSNs
in a solar cycle when the start time of the cycle has been determined.
\edit1{To predict the solar cycle nearly two cycles in advance,}
we obtain a new model denoted as TMLP-extension (hereafter referred to as TMLP-E)
by extending the prediction ability of TMLP
\edit1{in this work}.
If the start time of a cycle $n$ is already known, TMLP-E can predict the
variation of SSNs in the cycle $n+1$. In September 2020, the SILSO World Data
center confirmed that cycle 25 started in December 2019
(see http://sidc.be/silso/node/167/\#NewSolarActivity). Therefore, the
variations of SSNs in cycles 25 and 26 can be predicted by the TMLP and TMLP-E
models, respectively. The data used in this work are introduced in
Section~\ref{sec:data}. The TMLP and TMLP-E models are described in
Section~\ref{sec:model}. Prediction results of cycles 25 and 26 are reported
in Section~\ref{sec:result}. Conclusions and discussion are presented in
Section~\ref{sec:discussion}.

\section{Data}
\label{sec:data}
The Version 2.0 international SSN, issued by the Solar Influences Data Analysis
Center since 2015, is used in this study. The monthly SSN is available for all
cycles so that the information extracted from it can be used as potential
predictors to construct prediction models statistically. The monthly smoothed SSN,
obtained by using the standard smoothing with a time window of 13 months to smooth
the monthly SSN with half weights for the months at the start and end
\citep[e.g.,][]{Hathaway15}, is widely used to represent the solar activity so that
it will be predicted for cycles 25 and 26 in the following sections. In this work,
the monthly and monthly smoothed SSNs are denoted as $R$ and $S$, respectively.

\begin{figure}
\epsscale{1.15} \plotone{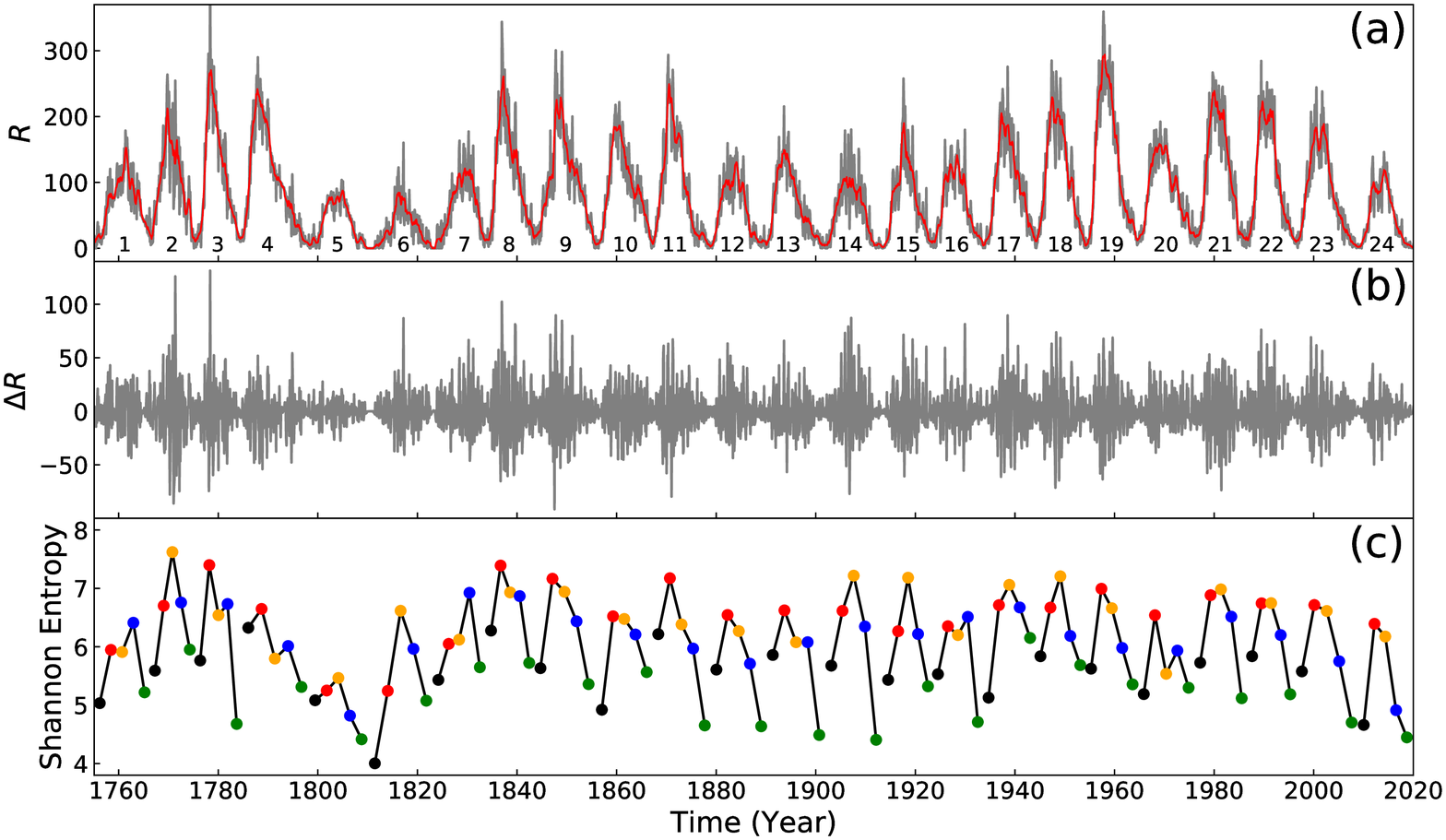}
\caption{(a) The gray curve shows the monthly SSN for cycles 1$-$24 with the
cycle numbers marked, while the red curve presents the running mean value of
monthly SSN with a time window of 9 months. (b) The fluctuation of monthly SSN
obtained by subtracting the running mean SSN from the monthly SSN.
(c) The Shannon entropy is presented with different colors for the 5 phases
of each cycle.}
\label{fig:entropy}
\end{figure}

The monthly SSN is utilized to obtain the Shannon entropy of solar cycles. The Shannon
entropy, also known as information entropy, was proposed by \citet{Shannon48} to
characterize the inherent randomness of a system quantitatively and applied to the
study of solar energetic particles \citep[e.g.,][]{LaurenzaEA12, QinAZhao13} as well as
solar activity \citep[e.g.,][]{KakadEA15, KakadEA17, KakadEA20, QinAWu18} in recent years.
The fluctuation of monthly SSN can be treated as a random system, and thereby the
Shannon entropy can be obtained from it. Figure~\ref{fig:entropy}(a) presents the
monthly SSN for cycles 1$-$24 as the gray curve, and the number in the figure
indicates the cycle number. We use a time window of 9 months to obtain the running
mean value of monthly SSN, which is plotted as the red curve in
Figure~\ref{fig:entropy}(a). Thus, the fluctuation of monthly SSN, $\Delta R$, can
be obtained by subtracting the running mean value from the monthly SSN with
\begin{equation}
\Delta R(i) = R(i) - \frac{1}{w} \sum_{k=i-(w-1)/2}^{i+(w-1)/2} R(k),
\end{equation}
where $w$ is the time window.
The time sequence of $\Delta R$ is presented in Figure~\ref{fig:entropy}(b). It is
shown that the variation of $\Delta R$, to some extent, is cyclical, so that we
divide each solar cycle to 5 phases evenly to calculate the Shannon entropy of every
phase for extracting the characteristics of solar cycles at different phases
\citep[e.g.,][]{KakadEA17, QinAWu18}. The Shannon entropy of each phase, which is
computed from the probability density function of $\Delta R$ using the histogram
method, can be written as \citep{Wallis06, KakadEA15, KakadEA17}
\begin{equation}
E = - \sum_{k=1}^{m} p_k \text{log}_2 (p_k) + \text{log}_2 (d),
\end{equation}
where $E$ is the Shannon entropy, $p_k$ is the probability of the $k$th bin of the
histogram, $m$ is the number of bins, and $d=3.49\sigma N^{-1/3}$ is the bin width
with $\sigma$ and $N$ being the standard deviation and sample size of $\Delta R$,
respectively \citep{Scott79}. The Shannon entropy is presented with different colors
for the 5 phases of each cycle in Figure~\ref{fig:entropy}(c). It is shown that the
Shannon entropy also shows cyclical characteristics. The values of Shannon entropy
are 4.7, 6.4, 6.2, 4.9, and 4.4 for the 5 phases of cycle 24, and that of cycles
1$-$23 can be found in Table 4 of \citet{QinAWu18}.

The monthly smoothed SSN is adopted to characterize the solar maximum $S_m$ and
minimum $S_0$ of solar cycles and further to quantify the cycle length $T_c$ and
ascent time $T_a$. The solar maximum is defined as the mathematical maximum of the
monthly smoothed SSN in each cycle, while the solar minimum is defined as the
mathematical minimum of the monthly smoothed SSN in the period from the preceding
solar maximum to the current one. Note that the first occurrence of maximum/minimum
value is chosen as the epoch of solar maximum/minimum if the maximum/minimum value
occurs more than once in the period. The cycle length are the time interval between
the two solar minima at the beginning and end of the solar cycle, while the ascent
time are the time interval between the solar minimum at the beginning of the solar
cycle and the solar maximum. Cycle 25 started in December 2019, so that the length
of cycle 24 is exactly 11 yr and the solar minimum of cycle 25 equals to 1.8.

\section{Prediction Model}
\label{sec:model}

\subsection{TMLP Model}
In \citet{QinAWu18}, the variation of SSN in a solar cycle is described by the
modified logistic function, i.e.,
\begin{eqnarray}
S(x)&=&r_0\left(1-\frac{x^\alpha}{x_m^\alpha}\right)x, \label{eq:S_x}\\
x\left(t\right)&=&x_m {\left[1+\left(\frac{x_m^\alpha}{x_0^\alpha}-1\right)
 e^{-\alpha r_0t}\right]^{-1/\alpha}},
\end{eqnarray}
where $x$ is the cumulative SSN, $t$ is the elapsed
time from the solar minimum in units of months, $r_0$ is the maximum emergence rate
of sunspots, $x_m$ is the maximum cumulative SSN or total SSN that can be generated
in the cycle, $x_0$ is the initial cumulative SSN, and $\alpha$ is the asymmetry of
the cycle shape. Hence, the cycle features $S_m$, $S_0$, $T_c$, and $T_a$ can be
expressed as
\begin{eqnarray}
S_m &=& \frac{\alpha}{1+\alpha}\left(\frac{1}{1+\alpha}\right)^{1/\alpha}r_0x_m,\\
S_0 &=& r_0\frac{x_0}{x_m}\frac{x_m^\alpha-x_0^\alpha}{x_m^{\alpha-1}}, \label{eq:S0}\\
T_c &=& \left(12\alpha r_0\right)^{-1}
 \ln{\frac{x_m^\alpha/x_0^\alpha-1}{x_m^\alpha/x_e^\alpha-1}},\\
T_a &=& \left(12\alpha r_0\right)^{-1}\ln{\frac{x_m^\alpha/x_0^\alpha-1}{\alpha}},
\end{eqnarray}
where $x_e$ is the actual total SSN observed in the cycle and can be expressed as
$x_e=0.9722x_m-6.93$. Note that the units of $T_c$ and $T_a$ are in years.
The parameters $r_0$ and $\alpha$ were set to 0.2 and 0.224, respectively,
for facilitating the construction of SSN prediction model, which was called the
TMLP model in \citet{QinAWu18}.

The TMLP model can predict the variation of SSN in a solar cycle at the start of the
cycle. The key of the prediction made by TMLP is to predict the values of $x_m$ and
$x_0$. The parameter $x_m$ of cycle $n$ can be predicted by using the Shannon entropy
in the three preceding cycles and the cycle length of the last cycle, i.e.,
\begin{eqnarray}
x_m^{\left(n\right)}= &-&3509 E_4^{\left(n-3\right)}+3097 E_2^{\left(n-2\right)}+4327 E_5^{\left(n-2\right)} -3190 E_1^{\left(n-1\right)}\nonumber\\
 &+& 3189 E_4^{\left(n-1\right)}+1397 E_5^{\left(n-1\right)}-624 T_c^{\left(n-1\right)}-11862,
\label{eq:xm}
\end{eqnarray}
where the superscript denotes the cycle number, and the subscript of Shannon entropy,
$E$, is the phase number of the cycle. The parameter $x_0^{\left(n\right)}$ can be
obtained approximately by solving Equation~(\ref{eq:S0}) with the predicted
$x_m^{\left(n\right)}$ and the observed solar minimum $S_0^{\left(n\right)}$.

The observation and prediction for cycles 4$-$24 are presented in
Figure~\ref{fig:prediction} with the gray solid and red dot-dashed lines,
respectively, and the absolute relative errors of the predicted cycle features are
listed in columns 2$-$4 of Table~\ref{tab:prediction}. Note that the TMLP model is
constructed with the statistical analysis on the observations of cycles 1$-$23
\citep{QinAWu18}. It is shown that the prediction fits the observation well.
The average absolute relative errors of the predicted $S_m$, $T_a$, and $T_c$ are
8.5\%, 15.8\%, and 9.0\% for cycles 4$-$24, respectively. The confidence intervals
(CIs) of the predicted $S_m$, $T_a$, and $T_c$ at a 95\% significance level are
(-22.7\%, 18.4\%), (-34.6\%, 38.8\%), and (-23.6\%, 17.3\%), respectively.

\subsection{Extending the Prediction Ability of TMLP Model}
\begin{figure}
\epsscale{1} \plotone{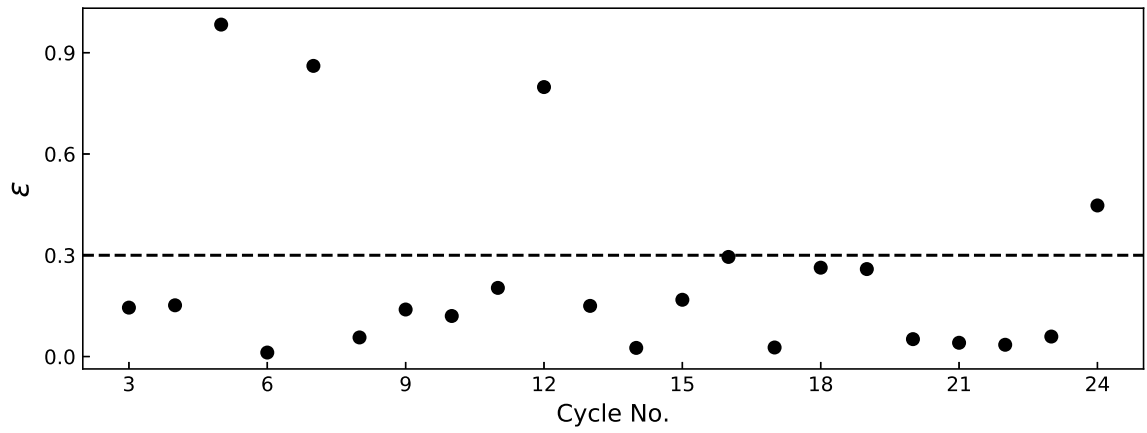}
\caption{The absolute relative error $\epsilon$ given by
Equation~(\ref{eq:epsilon}) for cycles 3$-$24.}
\label{fig:epsilon}
\end{figure}

Suppose it is currently in cycle $n$, to predict the solar activity of the next
cycle, cycle $n+1$, Equation~(\ref{eq:xm}) can be rewritten as
\begin{eqnarray}
x_m^{\left(n+1\right)}= &-&3509 E_4^{\left(n-2\right)}+3097 E_2^{\left(n-1\right)}+4327 E_5^{\left(n-1\right)} -3190 E_1^{\left(n\right)}\nonumber\\
 &+& 3189 E_4^{\left(n\right)}+1397 E_5^{\left(n\right)}-624 T_c^{\left(n\right)}-11862.
\label{eq:xm_temp}
\end{eqnarray}
At the start of cycle $n$, the parameters $E_1^{\left(n\right)}$,
$E_4^{\left(n\right)}$, $E_5^{\left(n\right)}$, and $T_c^{\left(n\right)}$ are unknown.
\edit1{We define the unknown terms as}
\begin{equation}
h(n) = -3190 E_1^{\left(n\right)} + 3189 E_4^{\left(n\right)} + 1397 E_5^{\left(n\right)}
  - 624 T_c^{\left(n\right)},
\end{equation}
\edit1{and we estimate $h(n)$ by assuming $h(n) = h(n-1)$, so that the absolute relative
error introduced by the assumption for predicting $x_m^{n+1}$ can be expressed as}
\begin{equation}
\epsilon(n+1) = \frac{\left| h(n) - h(n-1) \right|}{x_m^{\left(n+1\right)}}.
\label{eq:epsilon}
\end{equation}
\edit1{Note that the value of $x_m^{\left(n+1\right)}$ in Equation~(\ref{eq:epsilon})
is obtained by fitting Equation~(\ref{eq:S_x}) to the SSN of cycle $n+1$.
Figure~\ref{fig:epsilon} showed that the absolute relative error $\epsilon$ is not
greater than 30\% for most cycles, which indicates the assumption $h(n) = h(n-1)$
will not introduce large errors for most cycles.}
Therefore, we replace $E_i^{\left(n\right)}$ ($i=1, 4, 5$) with
$E_i^{\left(n-1\right)}$, and replace $T_c^{\left(n\right)}$ with
$T_c^{\left(n-1\right)}$ in Equation~(\ref{eq:xm_temp}).
Then, Equation~(\ref{eq:xm_temp}) can be written as
\begin{eqnarray}
x_m^{\left(n+1\right)}= &-&3509 E_4^{\left(n-2\right)} - 3190 E_1^{\left(n-1\right)}
  + 3097 E_2^{\left(n-1\right)} + 3189 E_4^{\left(n-1\right)} \nonumber\\
 &+& 5724 E_5^{\left(n-1\right)} - 624 T_c^{\left(n-1\right)} - 11862.
\label{eq:xm_new}
\end{eqnarray}
\edit1{To predict the value of $x_0^{\left(n+1\right)}$, we also assume
$S_0^{\left(n+1\right)}=S_0^{\left(n\right)}$ for simplicity, and thus}
$x_0^{\left(n+1\right)}$ can be obtained approximately by solving
Equation~(\ref{eq:S0}) with the predicted $x_m^{\left(n+1\right)}$ and the current
solar minimum $S_0^{\left(n\right)}$. All parameters are available for predicting
the variation of SSN in the cycle $n+1$ if the start time of the current cycle $n$
is already known. The new model for predicting the cycle $n+1$ is denoted as TMLP-E.

\subsection{Evaluating the Prediction Ability of TMLP-E Model}
\label{sec:evaluate}
\begin{figure}
\epsscale{1.15} \plotone{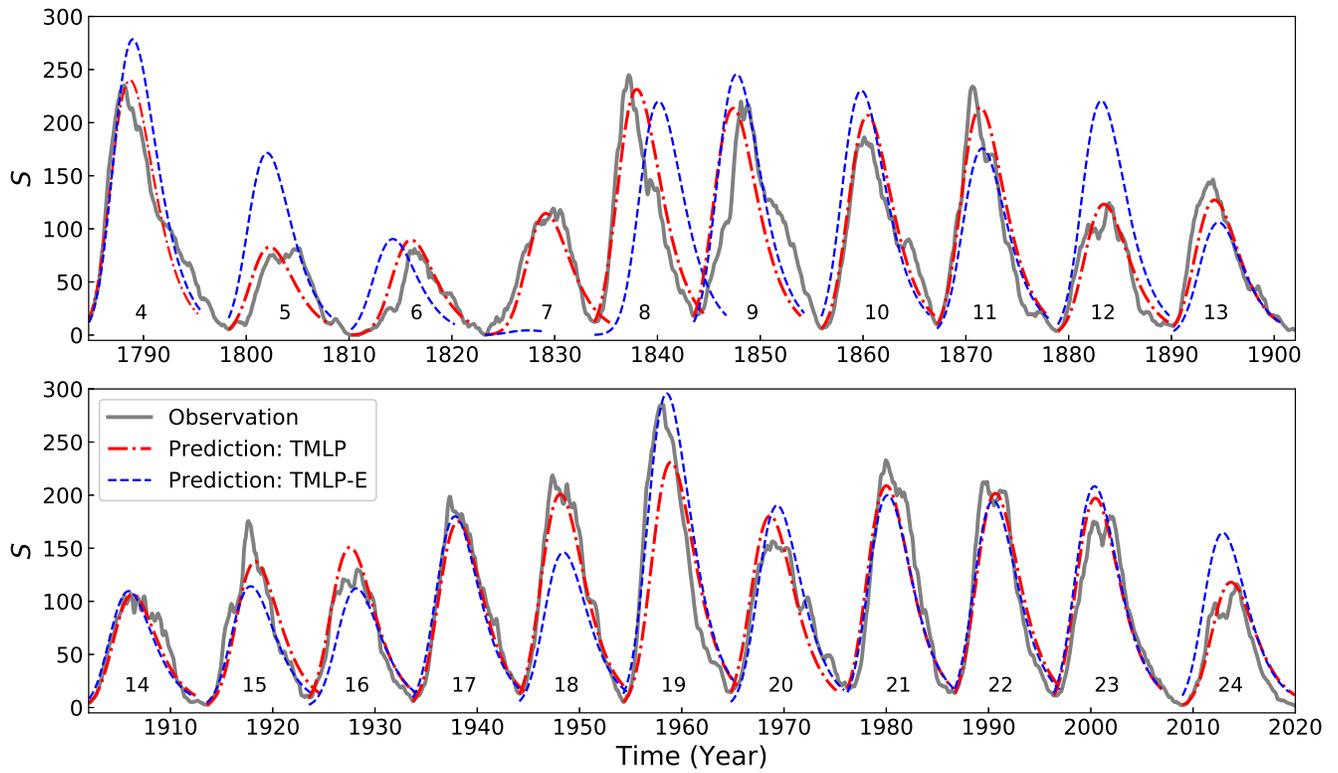}
\caption{The gray, red dot-dashed, blue dashed lines are the observed monthly
smoothed SSN, prediction by TMLP, and prediction by TMLP-E, respectively,
for cycles 4$-$24 with the cycle numbers marked.}
\label{fig:prediction}
\end{figure}

The prediction results of the TMLP-E for cycles 4$-$24 are presented in
Figure~\ref{fig:prediction} with the blue dashed lines, and the absolute relative
errors of the predicted cycle features are listed in columns 5$-$7 of
Table~\ref{tab:prediction}. Although $x_m$ and $x_0$ are obtained approximately,
the prediction result is, however, acceptable for most cycles.
\edit1{The average absolute relative errors of $S_m$, $T_a$, and $T_c$
are 29.0\%, 21.8\%, and 11.7\%, respectively.}
The predicted solar maximum deviates from the observation within 20\% for more than
half of the cycles, while cycles with errors no more than 35\% account for 81\%.

The prediction errors of solar maximum of cycles 5, 7, 12, and 24 are greater than
35\%. Cycles 5, 12, and 24 have a common feature that the solar maximum has a
sudden drop (more than 30\%) compared to the solar maximum of the preceding cycle.
Besides, the Sun entered minima of Gleissberg cycle since cycle 5 and cycle 12,
and the Sun might also enter a minimum of Gleissberg cycle since cycle 24. On the
other hand, cycle 20, which also has a solar maximum more than 30\% less than the
preceding one but is not during a minimum of Gleissberg cycle, is predicted with a
moderate error (21.5\%) for the solar maximum. Therefore, it is supposed that the model
can not predict the first solar cycle of the minimum of a Gleissberg cycle accurately.
For cycle 7, all the cycle features are predicted with large error, and
the prediction is indeed
\edit1{abnormal}
according to the predicted time profile as shown in Figure~\ref{fig:prediction}.
The prediction is made by mainly using the data of cycle 5 that is during the
Dalton minimum. The Shannon entropy of cycle 5 is at an unusually low level
because the Dalton minimum has unusually long periods of sunspot inactivity,
which might be the reason why the prediction is
\edit1{abnormal}.
Therefore, an
\edit1{abnormal}
prediction may imply a Dalton-like minimum.


\edit1{Based on the above analysis, if we could determine that the cycle to be
predicted will not be the first cycle of a new Gleissberg cycle and the predicted
result is not abnormal, cycles 5, 7, 12 and 24 can be excluded for evaluating the
prediction accuracy of the TMLP-E model. With cycles 5, 7, 12 and 24 excluded,
the average absolute relative errors of $S_m$, $T_a$, and $T_c$ reduce to 16.8\%,
19.7\%, and 10.3\%, respectively.}
The CIs of $S_m$, $T_a$, and $T_c$ are (-41.9\%, 33.9\%), (-43.1\%, 62.5\%), and
(-28.9\%, 22.8\%), respectively. It is shown that both the relative error and CI
of the predicted $S_m$ of TMLP-E are about twice that of TMLP. Besides, both the
relative error and CI of the predicted $T_c$ of TMLP-E are similar to that of TMLP.
In addition, the relative error and CI of the predicted $T_a$ of TMLP-E are about
25\% and 44\% larger than that of TMLP, respectively.

The solar maximum of the TMLP-E result (blue line) is either the largest or the
smallest one among the results of observations, TMLP, and TMLP-E in each cycle as
shown in Figure~\ref{fig:prediction} except for cycle 17. Therefore, the solar
maximum predicted by TMLP-E can be used as either the upper or lower limit to narrow
down the CI of the solar maximum predicted by TMLP, although it will slightly
increase the prediction error of cycle 17. The solar maxima of the results from
observations, TMLP, and TMLP-E for cycles 4$-$24 are listed in columns 2, 3, and 4
of Table~\ref{tab:CI}, respectively. Column 5 gives the CI of the predicted solar
maximum of TMLP. The modified CI, listed in Column 6, is obtained by choosing the
predicted solar maximum of TMLP-E as either the upper or lower limit if the
predicted solar maximum of TMLP-E is within the CI. Column 7 shows whether the
prediction of TMLP is improved if we use the modified CI. Note that cycles 5, 7, 12,
and 24 are not listed in Table~\ref{tab:CI}. It is shown that almost 82\% of CIs
predicted by TMLP can be modified, and 93\% of the modifications can narrow down
the CIs correctly.

\section{Results}
\label{sec:result}
\begin{figure}
\epsscale{0.85} \plotone{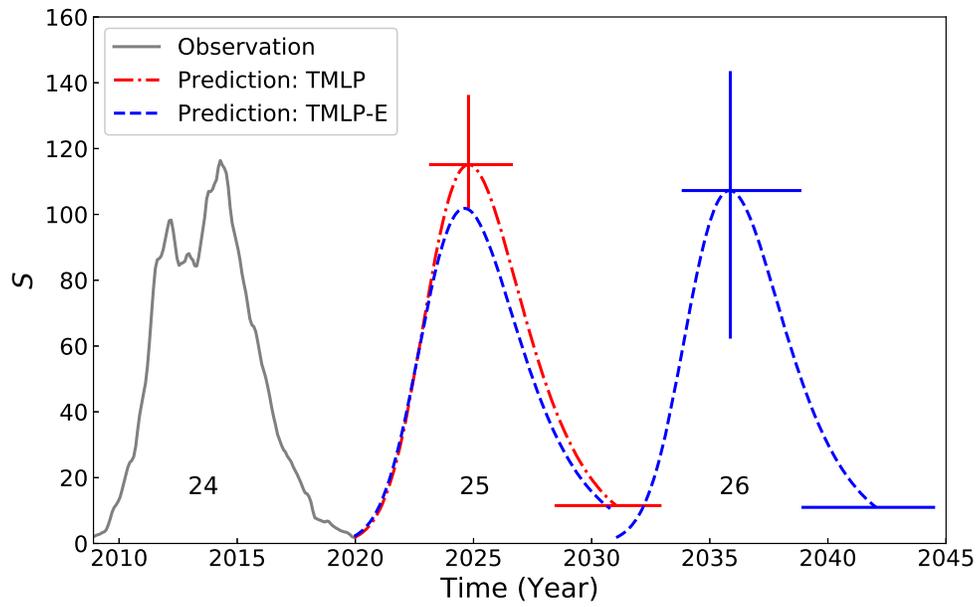}
\caption{Similar to Figure~\ref{fig:prediction} but for cycles 24$-$26.
The vertical error bar shows the CI of the cycle amplitude at a 95\%
significance level, and the horizontal error bars at the solar maximum and
minimum denote the CI of the peak and end time of the cycle, respectively.}
\label{fig:prediction_25_26}
\end{figure}

In Figure~\ref{fig:prediction_25_26}, the red dot-dashed line indicates the
predicted result of TMLP for cycle 25, and the blue dashed lines indicate the
results of TMLP-E for cycles 25 and 26.
\edit1{It is shown that the predicted time profiles of cycles 25 and 26 by the
TMLP-E are the typical rather than abnormal shape of the solar cycle. Besides,}
it is suggested that cycle 24 rather than cycle 25 may be the first cycle of a
minimum of Gleissberg cycle in Section~\ref{sec:evaluate}, so that
\edit1{cycles 25 and 26}
can be predicted with a good accuracy
\edit1{by the TMLP-E. The error bars in the figure denote the CIs at a 95\%
significance level in the amplitude, peak time and end time.}
Table~\ref{tab:prediction_25_26} gives the values of the predicted cycle features
and their CIs. For comparison, the observations of cycle 24 are also presented
in the figure and table.

Both models TMLP and TMLP-E could be used for the prediction of cycle 25, here,
we use the results from TMLP (red dot-dashed line) as the prediction of cycle 25,
and the predicted $S_m$, $T_a$, and $T_c$ are 115.1, 4.84 yr, and 11.06 yr,
respectively. In addition, the CIs of the predicted $S_m$, $T_a$, and $T_c$ are
(89.0, 136.3), (3.17 yr, 6.72 yr), and (8.45 yr, 12.98 yr), respectively.
On the other hand, if we use model TMLP-E to predict cycle 25, the $S_m$ is
obtained as 101.8, which is less than the $S_m$ predicted by TMLP, so that the
CI of the $S_m$ can be modified to (101.8, 136.3).

Model TMLP-E can be used to predict cycle 26 shown in Figure 
\ref{fig:prediction_25_26}, and the predicted $S_m$, $T_a$, and $T_c$ are 107.3,
4.80 yr, and 10.97 yr, respectively. Furthermore, the CIs of the $S_m$, $T_a$,
and $T_c$ are (62.3, 143.6), (2.74 yr, 7.81 yr), and (7.81 yr, 13.47 yr),
respectively. It is shown that the predictions of cycles 25 and 26 are similar
to the observation of cycle 24.

\section{CONCLUSIONS AND DISCUSSION}
\label{sec:discussion}

In this paper, we use the TMLP and TMLP-E models to predict the SSNs of cycles 25
and 26. The TMLP model, which is proposed by \citet{QinAWu18}, has two parameters,
namely, the maximum cumulative SSN $x_m$ and the initial cumulative SSN $x_0$.
The parameter $x_m$ of cycle $n$ can be expressed as the linear combination of the
Shannon entropy of the three preceding cycles and the length of the last cycle, i.e.,
Equation~(\ref{eq:xm}). The other parameter $x_0$ can be estimated by solving
Equation~(\ref{eq:S0}) with the predicted $x_m$ and observed $S_0$ of cycle $n$.
Therefore, the variation of SSNs in cycle $n$ can be predicted by the TMLP model if
the start time of cycle $n$ is already known. To predict the variation of SSNs in
cycle $n+1$, we extend the TMLP to TMLP-E by assuming that the behavior of cycle $n$
would be similar to that of cycle $n-1$, i.e., one may replace the Shannon entropy
and cycle length of cycle $n$ with that of cycle $n-1$, and replace the solar 
minimum of cycle $n+1$ with that of cycle $n$. With this assumption, $x_m$ of cycle
$n+1$ can be predicted with the Shannon entropy of cycles $n-1$ and $n-2$ and the
length of cycle $n-1$, and $x_0$ can be estimated with the predicted $x_m$ and the
solar minimum of cycle $n$. Therefore, the variation of SSN in cycle $n+1$ can be
predicted by the TMLP-E model when the start time of cycle $n$ has been determined.
All in all, the TMLP and TMLP-E models can predict monthly smoothed SSNs nearly one
and two cycles in advance, respectively.

The prediction ability of TMLP-E is evaluated with the prediction accuracy of
cycles 4$-$24. Note that, cycles 5, 7, 12, and 24 are excluded if we are not going
to predict the first cycle of the minimum of a Gleissberg cycle, and if we do not
consider the
\edit1{abnormal}
predictions. The average absolute relative errors of the solar maximum, ascent time,
and cycle length predicted by TMLP-E are 16.8\%, 19.7\%, and 10.3\%, respectively.
CIs of the predicted solar maximum, ascent time, and cycle length are
(-41.9\%, 33.9\%), (-43.1\%, 62.5\%), and (-28.9\%, 22.8\%), respectively, at a
95\% significance level. Besides, the solar maximum predicted by TMLP-E can be used
to narrow down the CI of the solar maximum predicted by TMLP for most cycles.

The solar maximum of cycle 25 predicted by TMLP is 115.1, and the CI at a 95\%
significance level is (89.0, 136.3), which can be further modified to (101.8, 136.3)
by the predicted solar maximum of TMLP-E. The solar maximum would occur in October
2024 (95\% CI is from February 2023 to September 2026), and cycle 25 would end in January
2031 (95\% CI is from May 2028 to December 2032). The solar maximum of cycle 26 predicted
by TMLP-E is 107.3 (95\% CI is 62.3$-$143.6). If the end time of cycle 25 predicted
by the TMLP is chosen as the start time of cycle 26, the solar maximum is expected to
appear in November 2035 (95\% CI is from October 2033 to November 2038), and cycle 26
would end in January 2041 (95\% CI is from November 2038 to July 2044). The solar maxima
of cycles 25 and 26 are predicted to be at a low level and similar to that of cycle
24, which is about 40\% greater than the solar maximum of cycle 5 or cycle 6.
We therefore suggest that the declining trend of solar activity will break and cycles
24$-$26 are at a minimum of Gleissberg cycle rather than a Dalton-like minimum.

The Solar Cycle Prediction Panel experts released a forecast for cycle 25 in 2019
(see https://www.weather.gov/news/190504-sun-activity-in-solar-cycle), and the solar
maximum was expected to be in the range between 95 and 130 and would peak during the
time interval 2023$-$2026.
\edit1{The mean predicted amplitude of four physical model based forecasts for
cycle 25 is $110.5\pm13.5$ \citep{Nandy21}. Since}
the prediction results for cycle 25 in this work are very similar to that reported by
the Solar Cycle Prediction Panel experts
\edit1{and that predicted by physics-based forecasts, we believe that the TMLP-E
model derived from the TMLP model can predict the cycle 26 with a good accuracy.}

Several research inferred that the strong suppression of some parameters such as the
occurrence rate of flares in cycle 23 compared to cycle 22 may be the earlier sign of
the sudden drop of solar activity from cycle 23 to 24
\citep[][and references therein]{Petrovay20}. In this work, TMLP-E can not predict
the first solar cycle of the minimum of Gleissberg cycle accurately about two cycles
in advance while TMLP can forecast it well about one cycles in advance. Therefore, we
suggest that the preceding cycle is important for predicting the sudden drop of solar
activity or the start of a new Gleissberg cycle.

\acknowledgments
This work was supported, in part, under grants NNSFC 41874206 and NNSFC 42074206.
We thank the SIDC and SILSO teams and the Royal Observatory of Belgium for
international sunspot data. Figures were prepared with Matplotlib \citep{Hunter07}.



\clearpage

\clearpage

\clearpage

\clearpage


\clearpage
\begin{deluxetable}{c rrr c rrr}
\tablecaption{Absolute relative errors of predicted cycle features for cycles 4$-$24.
\label{tab:prediction}}
\tablehead{
& \multicolumn{3}{c}{TMLP} & & \multicolumn{3}{c}{TMLP-E} \\
\cline{2-4}\cline{6-8}
\colhead{Cycle No.} & \colhead{$\delta S_m/S_m$} & \colhead{$\delta T_a/T_a$} &
\colhead{$\delta T_c /T_c$} & & \colhead{$\delta S_m/S_m$} &
\colhead{$\delta T_a/T_a$} & \colhead{$\delta T_c /T_c$}
}
\startdata
4 & 2.0\% & 16.3\% & 22.1\% & & 18.4\% & 24.4\% & 19.6\% \\
5 & 1.1\% & 41.5\% & 16.4\% & & 109.1\% & 45.3\% & 14.4\% \\
6 & 10.3\% & 5.4\% & 8.8\% & & 11.4\% & 34.2\% & 22.4\% \\
7 & 3.9\% & 10.8\% & 14.0\% & & 96.3\% & 36.8\% & 46.6\% \\
8 & 5.5\% & 23.7\% & 10.9\% & & 10.2\% & 87.4\% & 32.6\% \\
9 & 2.9\% & 16.6\% & 16.4\% & & 11.7\% & 9.2\% & 13.2\% \\
10 & 11.1\% & 7.8\% & 1.9\% & & 23.6\% & 7.0\% & 7.0\% \\
11 & 8.4\% & 23.2\% & 8.3\% & & 24.9\% & 28.6\% & 7.6\% \\
12 & 1.0\% & 10.6\% & 4.6\% & & 77.3\% & 15.5\% & 4.0\% \\
13 & 13.2\% & 3.9\% & 13.2\% & & 27.6\% & 14.3\% & 10.9\% \\
14 & 0.5\% & 4.4\% & 9.3\% & & 2.3\% & 4.9\% & 12.5\% \\
15 & 21.9\% & 16.6\% & 10.9\% & & 35.0\% & 5.5\% & 5.2\% \\
16 & 15.9\% & 15.5\% & 2.3\% & & 13.8\% & 2.2\% & 6.7\% \\
17 & 11.9\% & 23.1\% & 4.4\% & & 9.4\% & 15.2\% & 1.9\% \\
18 & 8.1\% & 22.9\% & 3.5\% & & 33.2\% & 32.4\% & 4.9\% \\
19 & 18.9\% & 18.7\% & 7.1\% & & 3.8\% & 8.4\% & 4.1\% \\
20 & 15.3\% & 5.7\% & 9.5\% & & 21.5\% & 11.1\% & 3.3\% \\
21 & 10.3\% & 1.3\% & 1.4\% & & 14.2\% & 4.6\% & 0.5\% \\
22 & 5.1\% & 25.3\% & 8.6\% & & 8.3\% & 18.4\% & 6.2\% \\
23 & 9.3\% & 25.9\% & 15.8\% & & 15.6\% & 27.5\% & 16.3\% \\
24 & 1.4\% & 11.0\% & 0.2\% & & 41.3\% & 25.8\% & 5.6\% \\
\hline
avg. & 8.5\% & 15.7\% & 9.0\% & & 29.0\% & 21.8\% & 11.7\% \\
std. & 6.2\% & 9.8\% & 5.8\% & & 29.6\% & 19.4\% & 11.2\%
\enddata
\end{deluxetable}

\clearpage
\begin{deluxetable}{c rrr cc c}
\tablecaption{Modification of CI of the solar maximum predicted by TMLP.
\label{tab:CI}}
\tablehead{
\colhead{Cycle No.} & \colhead{$S_m^{\text{obs}}$} & \colhead{$S_m^{\text{TMLP}}$}
 & \colhead{$S_m^{\text{TMLP-E}}$} & \colhead{CI} & \colhead{Modified CI}
 & \colhead{Improved}
}
\startdata
4 & 235.3 & 240.0 & 278.5 & (185.6, 284.1) & (185.6, 278.5) & Yes \\
6 & 81.2 & 89.6 & 90.4 & (69.3, 106.1) & (69.3, 90.4) & Yes \\
8 & 244.9 & 231.3 & 220.0 & (178.9, 273.8) & (220.0, 273.8) & Yes \\
9 & 219.9 & 213.5 & 245.7 & (165.1, 252.7) & (165.1, 245.7) & Yes \\
10 & 186.2 & 206.9 & 230.1 & (160.0, 244.9) & (160.0, 230.1) & Yes \\
11 & 234.0 & 214.3 & 175.7 & (165.7, 253.6) & (175.7, 253.6) & Yes \\
13 & 146.5 & 127.2 & 106.0 & (98.3, 150.5) & (106.0, 150.5) & Yes \\
14 & 107.1 & 106.6 & 109.6 & (82.4, 126.1) & (82.4, 109.6) & Yes \\
15 & 175.7 & 137.3 & 114.2 & (106.2, 162.5) & (114.2, 162.5) & Yes \\
16 & 130.2 & 150.9 & 112.2 & (116.7, 178.6) & --- & --- \\
17 & 198.6 & 175.0 & 180.0 & (135.3, 207.2) & (135.3, 180.0) & No \\
18 & 218.7 & 200.9 & 146.0 & (155.3, 237.8) & --- & --- \\
19 & 285.0 & 231.0 & 295.7 & (178.7, 273.5) & --- & --- \\
20 & 156.6 & 180.6 & 190.3 & (139.6, 213.8) & (139.6, 190.3) & Yes \\
21 & 232.9 & 208.8 & 199.9 & (161.5, 247.2) & (199.9, 247.2) & Yes \\
22 & 212.5 & 201.7 & 194.8 & (156.0, 238.7) & (194.8, 238.7) & Yes \\
23 & 180.3 & 197.2 & 208.3 & (152.5, 233.4) & (152.5, 208.3) & Yes
\enddata
\end{deluxetable}

\clearpage
\begin{deluxetable}{cc ccc cc cc}
\tablecaption{Predicted cycle features of cycles 25$-$26.
\label{tab:prediction_25_26}}
\tablehead{
\colhead{Cycle No.} & \colhead{Model} & \colhead{$S_m$} & \colhead{CI$_{S_m}$}
 & \colhead{Modified CI$_{S_m}$}
 & \colhead{$T_a$} & \colhead{CI$_{T_a}$} & \colhead{$T_{cy}$}
 & \colhead{CI$_{T_{cy}}$}
}
\startdata
24 & Observation & 116.4 & --- & --- & 5.33 & --- & 11.00 & --- \\
\multirow{2}{*}{25} & TMLP & 115.1 & (89.0, 136.3) & (101.8, 136.3) & 4.84
 & (3.17, 6.72) & 11.06 & (8.45, 12.98)\\
& TMLP-E & 101.8 & (59.2, 136.3) & --- & 4.66 & (2.65, 7.58) & 10.79 & (7.68, 13.25)\\
26 & TMLP-E & 107.3 & (62.3, 143.6) & --- & 4.80 & (2.74, 7.81) & 10.97 & (7.81, 13.47)
\enddata
\end{deluxetable}

\end{document}